\documentclass[10pt]{iopart}

\usepackage{graphicx}
\usepackage{exscale}
\usepackage{iopams}
\usepackage{subfigure}
\usepackage{xspace}
\usepackage{citesort}

\DeclareMathAlphabet{\bi}{OML}{cmm}{b}{it}
\newfont{\tensy}{cmsy10}
\newcommand{\chem}[1]{{$\fontdimen16\tensy=3.0pt
    \fontdimen17\tensy=3.0pt \mathrm{#1}$}}

\newcommand{\UP}[0]{\uparrow}
\newcommand{\DO}[0]{\downarrow}
\newcommand{\Ek}[0]{E_\text{kin}}
\newcommand{\om}[0]{\omega}
\newcommand{\si}[0]{\sigma}
\newcommand{\dtau}{\Delta\tau}
\newcommand{\ie}[0]{i.e.\@\xspace}
\newcommand{\eg}[0]{e.g.\@\xspace}
\newcommand{\op}{\hat{p}}
\newcommand{\ox}{\hat{x}}
\newcommand{\on}{\hat{n}}
\newcommand{\wf}{w^\text{f}}
\newcommand{\rD}{\text{D}}
\newcommand{\Ep}{E_\mathrm{P}}
\newcommand{\text}[1]{\mathrm{#1}}
\newcommand{\bm}[1]{\boldsymbol{#1}}
\renewcommand{\hat}[1]{\widehat{#1}}
\newcommand{\nag}{\phantom{\dag}}

\renewcommand{\tr}{\mathrm{Tr}}
\newcommand{\las}[0]{\langle}
\newcommand{\ras}[0]{\rangle}

\begin{document}

%%%%%%%%%%%%%%%%%%%%%%%%%%%%%%%%%%%%%%%%%%%%%%%%%%%%%%%%%%%%%%%%%%%%%%%%%%%%%
%%%%%%%%%%%%%%%%%%%%%       TITLE & ABSTRACT          %%%%%%%%%%%%%%%%%%%%%%%
%%%%%%%%%%%%%%%%%%%%%%%%%%%%%%%%%%%%%%%%%%%%%%%%%%%%%%%%%%%%%%%%%%%%%%%%%%%%%
\title[Bipolaron formation in quantum dots]%
{Bipolaron formation in 1D-3D quantum dots:\\
 A lattice quantum Monte Carlo approach}

\author{M Hohenadler$^1$ and H Fehske$^2$}

\address{$^1$%
  Theory of Condensed Matter, Cavendish Laboratory, Cambridge, United Kingdom
}
\address{$^2$% 
  Institute for Physics, Ernst-Moritz-Arndt University Greifswald, Germany
}

\ead{\mailto{mh507@cam.ac.uk}}

\begin{abstract}
  Polaron and bipolaron formation in the Holstein-Hubbard model with harmonic
  confinement potential, relevant to quantum dot structures, is investigated
  in one to three dimensions by means of unbiased quantum Monte Carlo
  simulations. The discrete nature of the lattice and quantum phonon effects
  are fully taken into account. The dependence on phonon frequency, Coulomb
  repulsion, confinement strength (dot size) and electron-phonon interaction
  strength is studied over a wide range of parameter values. Confinement is
  found to reduce the size of (bi-)polarons at a given coupling
  strength, to reduce the critical coupling for small-(bi-)polaron formation,
  to increase the polaron binding energy, and to be more important in lower
  dimensions. The present method also permits to consider models with
  dispersive phonons, anharmonic confinement, or long-range interactions.
\end{abstract}


% 71.38.-k : Polarons and electron-phonon interactions
% 71.38.Ht : Self-trapped or small polarons
% 71.38.Mx : Bipolarons 
% 61.46.-w : Nanoscale materials  (for electronic transport in nanoscale materials, see 73.63.-b)
% 63.22.+m : Phonons or vibrational states in low-dimensional structures and nanoscale materials 
% 73.21.La : Quantum dots 

%
%
%
%%%%%%%%%%%%%%%%%%%%%%%%%%%%%%%%%%%%%%%%%%%%%%%%%%%%%%%%%%%%%%%%%%%%
\section{Introduction}\label{sec:intro}
%%%%%%%%%%%%%%%%%%%%%%%%%%%%%%%%%%%%%%%%%%%%%%%%%%%%%%%%%%%%%%%%%%%%%
%
%
%

Continuous advances in fabrication methods in semi-conductor physics over
recent decades have enabled researchers to create nanoscale systems in which
carrier motion is confined along one or more spatial dimensions. Among these
structures are quantum dots \cite{quantumdots,Yoffe01}, corresponding to
quasi-zero-dimensional systems with some atom-like properties \cite{KoMa98},
which are of substantial technical interest due to their potential use as
lasers \cite{Akiyama98}, in quantum computing \cite{LiWuStGaStKaPaPiSh03}, as
storage devices \cite{KrDuHeBiScAbFi04}, or as single-photon sources
\cite{MiKiBeScPeZhHuIm00,PeSaVuZhSo02}. Improvements in experimental
techniques now also permit studies of the properties of individual dots
rather than ensembles \cite{FiHuLoRaHe01}, as well as of molecular quantum
dots \cite{LiShBoLoPa02}.

Apart from strong correlations between electrons, the interaction of charge
carriers with the lattice degrees of freedom is of great relevance,
especially concerning its role in dephasing and relaxation processes
\cite{InSa92,Pazy02,MuZi04} as well as transport \cite{WuCa04}. Evidence for
the existence of polarons (carriers bound to a self-induced lattice
distortion) has been given \cite{HaGuVeFeBa99}, and non-trivial multi-phonon
effects and polaron formation have been proposed to explain the lack of
experimental evidence for the phonon bottleneck in quantum dots
\cite{InSa92,PrGrFedVGuTePoLe06}.  Formation of bipolarons (bound electron
pairs sharing a phonon cloud) has been proposed as an explanation for pair
tunneling from \chem{GaAs} quantum dots \cite{WaOrPh97}. Besides,
(small) bipolaron states significantly influence the transport through single
molecules \cite{AlBr03,KoRavO06}.

The problem of polaron formation in a quantum dot has received a lot of
attention in the past (see \cite{YiEr91,YiEr91_2,Sa96,PoFoDeBaKl99,ChMu01}
and references therein). From this work, the general conclusion is that
confinement enhances the tendency of an electron to undergo a crossover to a
(small) polaron so that heavy polaronic quasiparticles may be realised
experimentally even in the weak or intermediate electron-phonon coupling
regime. Less work has been devoted to understand bipolaron formation in
quantum dots \cite{MuCh96,WaOrPh97,PoFoDeBaKl99,SeEr00,ChMu01}, with
contradictory results on the effect of confinement
\cite{PoFoDeBaKl99,SeEr00}.

All existing work is based on the continuum Fr\"ohlich model \cite{Fr54},
employing mainly variational approaches of different reliability---the
above-mentioned conflicting results seem to originate from the inadequacy of
some of the approaches used.  However, from a theoretical point of view, the
most interesting case is that in which the extension of the lattice
displacement attached to the charge carriers is comparable to the confinement
length. The Holstein molecular-crystal model has been originally developed
\cite{Ho59a} to answer the question as to whether a local lattice instability
occurs upon increasing the electron-phonon coupling. This problem cannot be
addressed in the framework of continuum models, since the local lattice
dynamics on the scale of the unit cell has to be taken into account
\cite{Ra06}.

The need for lattice models also stems from the fact that it is nowadays
possible to fabricate self-assembled quantum dots with lateral dimensions of
less than 4 nm \cite{FiHuLoRaHe01}, which contain only a small number of unit
cells in each direction. Obviously, in such systems, the discrete nature of
the lattice will play an important role, motivating us to revisit the problem
of (bi-)polaron formation.

The Holstein-Hubbard model considered here is not completely realistic, as it
only includes local electron-phonon and electron-electron interactions as
well as a coupling to dispersionless optical phonons. However, we are
interested in fundamental effects arising from the combination of
electron-phonon interaction and confinement in discrete systems. Besides,
many aspects of quantum dots may be understood using effective models
\cite{quantumdots}.

From numerous studies of Holstein models with one (two) electrons on a
discrete lattice \cite{AlMo95,FeAlHoWe06}, it is well known that the
non-linear process of (bi-)polaron formation represents a complex many-body
problem that may not be satisfactorily described by means of variational or
perturbative approaches.  The necessity for a universal all-coupling approach
to describe intermediate-coupling materials such as \chem{CdTe} has been
pointed out long ago \cite{YiEr91_2}, motivating the application of unbiased
numerical methods.

In this work, we extend the world-line quantum Monte Carlo (QMC) method of
\cite{dRLa82,dRLa83,dRLa84,deRaLa86} to include a harmonic confining
potential. The method yields practically exact and unbiased results in any
dimension, in principle with no restrictions of system size or parameters. As a
consequence, we will be able to study (bi-)polaron formation over the whole
range from weak to strong confinement.  The method is capable of treating
systems with long-range interactions or dispersive phonons, which opens up
interesting perspectives for future work. An important finding as compared to
previous work (including \cite{Ko97}) concerns the problem of very long
autocorrelations times in certain parameter regimes. Finally, in contrast to earlier
Green-function QMC calculations for a continuum model \cite{WaOrPh97}, the
present approach does not rely on a trial wavefunction.

The paper is organised as follows. In section~\ref{sec:model}, we present the
model and briefly discuss the underlying approximations. The QMC method is
described in section~\ref{sec:method}, and section~\ref{sec:results} contains
a discussion of our results. We end with a summary and an outlook to future
work in section~\ref{sec:summary}.

%
%
%%%%%%%%%%%%%%%%%%%%%%%%%%%%%%%%%%%%%%%%%%%%%%%%%%%%%%%%%%%%%%%%%%%%%
\section{Model}\label{sec:model}
%%%%%%%%%%%%%%%%%%%%%%%%%%%%%%%%%%%%%%%%%%%%%%%%%%%%%%%%%%%%%%%%%%%%%
%
%
%

We consider a Holstein-Hubbard model, supplemented by a harmonic confining
potential. Using dimensionless phonon variables, the Hamiltonian reads
\begin{eqnarray}\label{eq:hamiltonian}
  \fl
  H
  = 
  \underbrace{-t\sum_{\las i,j\ras \si}c^\dag_{i\si}
    c^{\nag}_{j\si}}_{H_\text{kin}}
  +
  \underbrace{U\sum_{\phantom{\las}i\phantom{\ras}} \on_{i\UP}\on_{i\DO}}_{H_\text{ee}}
  +
  \underbrace{K\sum_{\phantom{\las}i\phantom{\ras}} |\bm{r}_i|^2 \on_i}_{H_\text{con}}
  %\\
  +
  \underbrace{\frac{\om_0}{2}\sum_i \left(\op_i^2 + \ox_i^2 \right)}_{H_\text{ph} = H_\text{ph}^p + H_\text{ph}^x}
  \underbrace{-\alpha\sum_i \ox_i \on_i}_{H_\text{ep}}
  \,.
\end{eqnarray}
Here $c^\dag_{i\si}$ creates an electron of spin $\si$ on lattice site $i$,
and $H_\text{kin}$ describes the hopping of electrons between
nearest-neighbouring lattice sites $\las i,j\ras$ with hopping integral $t$.
The second term, $H_\text{ee}$, accounts for on-site Coulomb repulsion
between electrons, where $\on_{i\si}=c^\dag_{i\si}c^{\nag}_{i\si}$. The
harmonic confinement potential, its strength being measured by $K$, is given
by $H_\text{con}$, $\bm{r}_i$ denotes the position vector of an electron at
site $i$, and $\on_i = \sum_\si \on_{i\si}$. $H_\text{ph}$ constitutes the
kinetic ($H^p_\text{ph}$) and potential energy ($H^x_\text{ph}$)
of the lattice degrees of freedom, corresponding
to independent harmonic oscillators with energy $\om_0$.  Finally,
$H_\text{ep}$ mediates the coupling between the local electron occupation
$\on_i$ and the lattice distortion $\ox_i$ with coupling strength $\alpha$.
We use units such that $\hbar=k_\text{B}=e=1$, consider D-dimensional
hypercubic lattices of linear size $N$ and volume $N^\rD$, and assume
periodic boundary conditions in real space.

In terms of the coupling parameter $\alpha$, the atomic-limit ($t=0$) polaron
binding energy $\Ep$ is given by $\Ep=\alpha^2/(2\om_0)$.  We define the
usual dimensionless coupling constant $\lambda = 2\Ep/W$, with the free
bandwidth $W = 4 t \rD$. Furthermore, we introduce the adiabaticity ratio
$\gamma=\om_0/t$, which permits to distinguish between the adiabatic
($\gamma<1$) and the non-adiabatic ($\gamma>1$) regimes.

This paper is exclusively concerned with the cases of either one electron, or
two electrons of opposite spin forming a spin singlet. Obviously, for
$N_\text{e}=1$, $H_\text{ee}=0$ and spin indices may be dropped. In
experiments, the maximum number of electrons in a quantum dot is strongly
influenced by the dot size (set here by $K$), so that the low-density regime
considered here may be relevant in particular for small dots, which
are of interest due to their potential use as quantum bits
\cite{KrDuHeBiScAbFi04}. One and two-electron states also play an important
role in molecular transistors, corresponding (in the simplest case) to a
quantum dot with a single electronic level coupled to a vibrational mode
\cite{AlBr03,KoRavO06}.

As we are interested in fundamental effects due to confinement, the
Holstein-Hubbard-type model seems a good starting point, especially as a lot
of knowledge is available for $K=0$
\cite{deRaLa86,WeFeWeBi00,BoKaTr00,Mac04,HovdL05}. Although quantitative
results for realistic quantum dot systems are beyond the scope of this work,
let us briefly discuss the simplifications inherent to the
Hamiltonian~(\ref{eq:hamiltonian}).

Using a one-band model we ignore the existence of a band gap in
semi-conducting host materials. Hence, our considerations are restricted to
one or two carriers in the same (either conduction or valence) band. A
symmetric, parabolic confining potential---which can be realised
experimentally \cite{quantumdots}---is assumed for simplicity, although more
general situations may be studied.

We restrict ourselves to a simple effective electron-phonon coupling to a
dispersionless (optical) phonon branch. For such local phonon modes, and in
the absence of sharp interfaces, the bulk-phonon approximation is expected to
be a good starting point. The coupling to optical phonons has been shown to
dominate bipolaron formation \cite{WaOrPh97}.

The Hamiltonian~(\ref{eq:hamiltonian}) only contains on-site interactions,
which are not completely justified in the case of semi-conducting materials.
Long-range interactions will be the subject of future investigations.

%
%
%%%%%%%%%%%%%%%%%%%%%%%%%%%%%%%%%%%%%%%%%%%%%%%%%%%%%%%%%%%%%%%%%%%%%
\section{Method}\label{sec:method}
%%%%%%%%%%%%%%%%%%%%%%%%%%%%%%%%%%%%%%%%%%%%%%%%%%%%%%%%%%%%%%%%%%%%%
%
%
%

The QMC method used here is a straight-forward extension of
\cite{dRLa82,dRLa83,deRaLa86}. The appealing features of this approach are
the analytical integration over the phonon degrees of freedom---enabling us
to study the adiabatic regime---and the fact that the numerical effort is
essentially independent of system size. Besides, the formalism treats the
cases of one and two electrons on the same footing, and is general enough so
as to permit future studies of models with, \eg, dispersive phonons or
long-range interactions.

\subsection{Partition function}

The derivation of the relevant fermionic contribution $Z^\text{f}$ (the
bosonic part can be calculated exactly) to the partition function
$Z=\tr\,\rme^{-\beta H}$ is almost identical to \cite{dRLa83,deRaLa86}.
Dividing the imaginary-time axis $[0,\beta]$ ($\beta=(k_\text{B}T)^{-1}$ is
the inverse temperature) into intervals of length $\dtau=\beta/L\ll1$ we may
write
\begin{equation}\label{eq:suzuki-trotter}
  Z_L 
  \approx
  \tr \left[
    \rme^{-\dtau (H_\text{kin}+H_\text{con})}\rme^{-\dtau (H_\text{ph}^x+H_\text{ep})}\rme^{-\dtau H_\text{ph}^p} 
  \right]^L
  \,.
\end{equation}
The result for the partition function reads \cite{dRLa83}
\begin{equation}\label{eq:Z2}
  Z^\text{f}_L
  =
  \sum_{\{\bm{r}^{(\xi)}_\tau\}}
  \wf(\{\bm{r}^{(\xi)}_\tau\})
  \,,
\end{equation}
with the fermionic weight
\begin{eqnarray}
  \fl
  \wf(\{\bm{r}^{(\xi)}_\tau\})
  = 
  \exp
  \left\{\sum_{\tau,\tau'=1}^L F(\tau-\tau') 
    \sum_{\xi,\xi'=1}^{N_\text{e}}
    \delta_{\bm{r}_{\tau}^{(\xi)},\bm{r}_{\tau'}^{(\xi')}}
  \right\}
  \\\nonumber
  \hspace*{-1em}
  \times
  \exp
  \left\{
    -\dtau\sum_{\tau=1}^L\left[
      U \delta_{\bm{r}^{(1)}_\tau,\bm{r}^{(2)}_\tau}
      +
      K\sum_{\xi=1}^{N_\text{e}} |\bm{r}^{(\xi)}_\tau|^2
    \right]
  \right\}
  \prod_{\tau=1}^L
  \prod_{\xi=1}^{N_\text{e}}
  I(\bm{r}^{(\xi)}_{\tau+1}-\bm{r}^{(\xi)}_\tau)
  \,.
\end{eqnarray}

Here $N_\text{e}=1$ (2) for the polaron (bipolaron) problem, and the components of
the position vector of electron $\xi$ on time slice $\tau$,
$\bm{r}^{(\xi)}_\tau$, are denoted as $r^{(\xi)}_{\tau,\mu}$
($\mu=1,\dots,\rD$).  The fermion world-lines are subject to periodic
boundary conditions both in real space and imaginary time, and the sum in
equation~(\ref{eq:Z2}) is over all allowed configurations.

The retarded electron (self-)interaction due to electron-phonon coupling is
described by the memory function
\begin{equation}
  F(\tau) 
  =
  \frac{\om_0\dtau^3\alpha^2}{4L}
  \sum_{\nu=0}^{L-1} \frac{\cos(2\pi \tau\nu/L)}
  {
    1 - \cos(2\pi\nu/L) + (\om_0\dtau)^2/2
  }
  \,,
\end{equation}
and electron hopping contributes
\begin{equation}
  I(\bm{r})
  =
  \frac{1}{N^{\rD}}
  \sum_{\bm{k}}
  \cos (\bm{k}\cdot\bm{r})\,
  \rme^{2\dtau t \sum_\mu \cos k_\mu}
  \,.
\end{equation}

\subsection{Observables}

We define the expectation value of a static observable $O$ as
\begin{equation}
  \las O\ras 
  =
  Z^{-1}\tr\,\hat{O}\rme^{-\beta H}
  = 
  Z^{-1}_L \sum_{\{\bm{r}^{(\xi)}_\tau\}} O(\{\bm{r}^{(\xi)}_\tau\})
    \wf(\{\bm{r}^{(\xi)}_\tau\})
    \,.
\end{equation}

The fermion contribution to the total energy is obtained from
\begin{equation}
  %\fl
  E^\text{f}
  =
  -\frac{\partial}{\partial\beta} \ln Z^\text{f}_L
  = E^\text{f}_\text{kin} + E^\text{f}_\text{ep} + E^\text{f}_\text{ee} + E^\text{f}_\text{con}
\end{equation}
with the kinetic energy ($\bm{\delta}$ runs over all nearest-neighbour sites)
\begin{equation}
  E^\text{f}_\text{kin}
  =
  -\frac{t}{L} \sum_{\tau=1}^L 
  \sum_{\xi=1}^{N_\text{e}}
  \sum_{\bm{\delta}=\text{n.n.}}
  \left\langle
  \frac{I(\bm{r}^{(\xi)}_{\tau+1}-\bm{r}^{(\xi)}_\tau+\bm{\delta})}
  {I(\bm{r}^{(\xi)}_{\tau+1}-\bm{r}^{(\xi)}_\tau)}
  \right\rangle
\end{equation}
and the interaction energies
\begin{equation}
  E^\text{f}_\text{ep}
  =
  -\frac{1}{L} 
  \sum_{\tau,\tau'=1}^L
  \frac{\partial  F(\tau-\tau') }{\partial\dtau}
  \sum_{\xi,\xi'=1}^{N_\text{e}}
  \left\langle
    \delta_{\bm{r}_{\tau}^{(\xi)},\bm{r}_{\tau'}^{(\xi')}} 
  \right\rangle
  \,,
\end{equation}
\begin{equation}
  E^\text{f}_\text{ee}
  =
  \frac{1}{L} 
  U
  \sum_{\tau=1}^L
  \left\langle
  \delta_{\bm{r}^{(1)}_\tau,\bm{r}^{(2)}_\tau}
  \right\rangle
  \,,
\end{equation}
and
\begin{equation}
  E^\text{f}_\text{con}
  =
  \frac{1}{L} 
  K
  \sum_{\tau=1}^L
  \sum_{\xi=1}^{N_\text{e}}
  \left\langle
    |\bm{r}^{(\xi)}_\tau|^2
  \right\rangle
\,.
\end{equation}

Also of interest is the electron-lattice correlation function
measured in the direction $\mu=1$, 
\begin{equation}
  C_\text{ep}(r)
  =
  (\Ep\beta N_\text{e})^{-1}
  \sum_{i=1}^N
  \las
  \on_i \ox_{i+r}
  \ras
  \,,
  \quad
  r = 0,1,\dots,N-1
  \,,
\end{equation}
which fulfills the sum rule $\sum_r C_\text{ep}(r) = 1$ (for
$\Ep>0$). Within QMC, we have \cite{dRLa84} 
\begin{equation}
  C_\text{ep}(r)
  = 
  (\Ep\beta N_\text{e})^{-1}
  \sum_{\tau,\tau'=1}^L F(\tau-\tau')
  \sum_{\xi=1}^{N_\text{e}}
  \las
  \delta_{r^{(\xi)}_{\tau,1},r^{(\xi)}_{\tau',1}}
  \ras
  \,.
\end{equation}
Similarly, for $N_\text{e}=2$, we calculate the electron-electron correlation function
\begin{equation}
  C_\text{ee}(r)
  =
  \sum_{i=1}^N \las \on_{i\UP} \on_{i+r\DO} \ras
   \,,
  \quad
  r = 0,1,\dots,N-1
  \,,
\end{equation}
fulfilling $\sum_r C_\text{ee}(r)=1$, for which the QMC estimator reads
\begin{equation}
  C_\text{ee}(r)
   =
  \frac{1}{L} 
  \sum_{\tau=1}^L
  \left\langle
  \delta_{|r^{(1)}_{\tau,1}-r^{(2)}_{\tau,1}|,r}
  \right\rangle
   \,.
\end{equation}

\subsection{Simulation details}

The system described by the partition function~(\ref{eq:Z2}) is characterised
by an additional dimension (imaginary time), as well as by a complicated
retarded interaction. As first shown in \cite{dRLa82}, it may be simulated by
means of Markov Chain Monte Carlo in combination with the Metropolis-Hastings
algorithm \cite{MeRoTeTe53}. To improve convergence for critical parameters,
we use both local updates (change of a random component of the position
vector on a random time slice of one particle by one lattice site) and global
updates (translation of an entire world line by one lattice site)
\cite{deRaLa86}.

An important point overlooked in previous work
\cite{dRLa82,dRLa83,dRLa84,Ko97} is the problem of significant statistical
correlations between successive configurations due to the local updating.
Whereas previous authors performed measurements every $L$ steps, we find that
this is by far not sufficient to ensure statistically independent
measurements, especially at intermediate electron-phonon coupling. The
integrated autocorrelation time in such cases may well exceed tens of
thousand of MC steps, so that a careful binning analysis \cite{HGE03} is
required for every run to ensure correct results and error bars. Details on
this problem will be reported elsewhere. Although autocorrelation times are
expected to be generally shorter in continuous-time simulations (see, \eg,
\cite{Ko98}), the problem must not be neglected. A QMC algorithm entirely
free of autocorrelations has been presented in \cite{HoEvvdL03,HovdL05}, but
this approach suffers from restrictions in system size for $\rD>1$
\cite{HoEvvdL05}. Another problem of world-line algorithms is that the
acceptance rate for local updates approaches zero in the strong-coupling
regime. For $K/t\gtrsim1$ and $\lambda\gtrsim1$, this becomes also true for
global updates. Finally, we also detected convergence problems in the
vicinity of the small-(bi-)polaron crossover due to critical slowing down.

The only systematic error in the present calculations comes from the
Suzuki-Trotter approximation (equation~(\ref{eq:suzuki-trotter})). It may be
eliminated by working in continuous imaginary time \cite{Ko98}, or by
performing simulations at different $\dtau$ and exploiting the
$\dtau^2$-dependence of the results to scale to $\dtau=0$. To save some
computer time, here we have simply chosen a single (small) value of
$\dtau=0.05$, which ensures that the Suzuki-Trotter error is satisfactorily
small.

%
%
%
%%%%%%%%%%%%%%%%%%%%%%%%%%%%%%%%%%%%%%%%%%%%%%%%%%%%%%%%%%%%%%%%%%%%%
\section{Results}\label{sec:results}
%%%%%%%%%%%%%%%%%%%%%%%%%%%%%%%%%%%%%%%%%%%%%%%%%%%%%%%%%%%%%%%%%%%%%
%
%
%

We discuss our numerical results in relation to the available knowledge about
the Holstein-Hubbard model (equation~(\ref{eq:hamiltonian}) with $K=0$).
Similar to the non-confined case, the adiabaticity ratio $\gamma$ turns out
to have an important effect on the physics, and a comparison of the cases
$\gamma\ll1$ and $\gamma\gg1$ will be given for the 1D polaron. In higher
dimensions, we shall focus on the adiabatic regime $\gamma\ll1$, as the
low-lying optical phonon energies in quantum dot materials are in the range
of 10-100 meV, \ie, substantially smaller than the electron bandwidth.

The 2D (3D) geometry considered here---corresponding to a circular
(spherical) quantum dot---may actually be realised experimentally, whereas
the 1D case is mainly of theoretical interest. However, we shall see that the
physics is qualitatively similar in $\rD=1$ and $\rD>1$.

All simulations have been carried out using a linear lattice size of $N=31$,
enough to obtain very well converged results even for $K=0$
\cite{dRLa83,Ko97}. Since for $K>0$ local physics becomes more important,
finite-size effects are negligible. The center of the parabolic potential is
located at site $i=0$ of each dimension, and the site indices range over
$[-15,15]$. To study ground-state properties, we set the inverse temperature
$\beta t=15$. The number of Trotter time slices has been fixed to $L=300$ so
that $\dtau=0.05$. Errorbars are typically smaller than the linewidth, and
are shown only if larger than the symbols used.

A drawback of the present approach (cf \cite{Ko98,PrSv98}) is that we can not
calculate the quasiparticle effective mass. Instead, we shall consider the
electronic kinetic energy, which includes contributions from incoherent
processes.  Additionally, the correlation functions defined in
section~\ref{sec:method} permit us to monitor the size of the (bi-)polaron,
and to determine the critical coupling for the crossover to a
small-(bi-)polaron state.

The confinement length---often a directly tunable parameter in previous
work---is set by the oscillator strength ($\sim K$) of the harmonic
potential. For two electrons, it may be estimated from the electron-electron
correlation function $C_\text{ee}(r)$ at $\lambda=0$.

\subsection{Polaron}

Existing work suggests that the ground state of the Holstein model with one
electron ($K=0$, $U$ irrelevant) at weak electron-phonon coupling is a large
polaron (extending over more than one lattice site) in 1D, and a quasi-free
electron for $\rD>1$ \cite{KaMa93}.

With increasing coupling $\lambda$, the potential energy due to lattice
deformation increases relative to the kinetic energy of the electron. In the
adiabatic regime $\gamma<1$, a small-polaron state (with the electron and the
lattice distortion being localised essentially to the same site) is formed if
$\Ep>W/2$ ($\lambda>1$). In contrast, for $\gamma>1$, small-polaron formation
requires a sizeable lattice distortion, leading to the condition
$g^2=\Ep/\om_0>1$. Note that these ``critical couplings'' are conceptually
different from those for the occurrence of the Peierls quantum phase
transition at half filling \cite{HoWeBiAlFe06} since there is no phase
transition in the (bi-)polaron problem for $\gamma>0$ \cite{Loe88}.

The small-polaron state is characterised by a substantially enhanced
effective mass and corresponding small quasiparticle weight. Generally, the size
of the polaron is smaller in the non-adiabatic strong-coupling regime due to
the faster lattice response.

\subsubsection{One dimension}

\begin{figure}
  \begin{center}
  \includegraphics[height=0.34\textwidth]{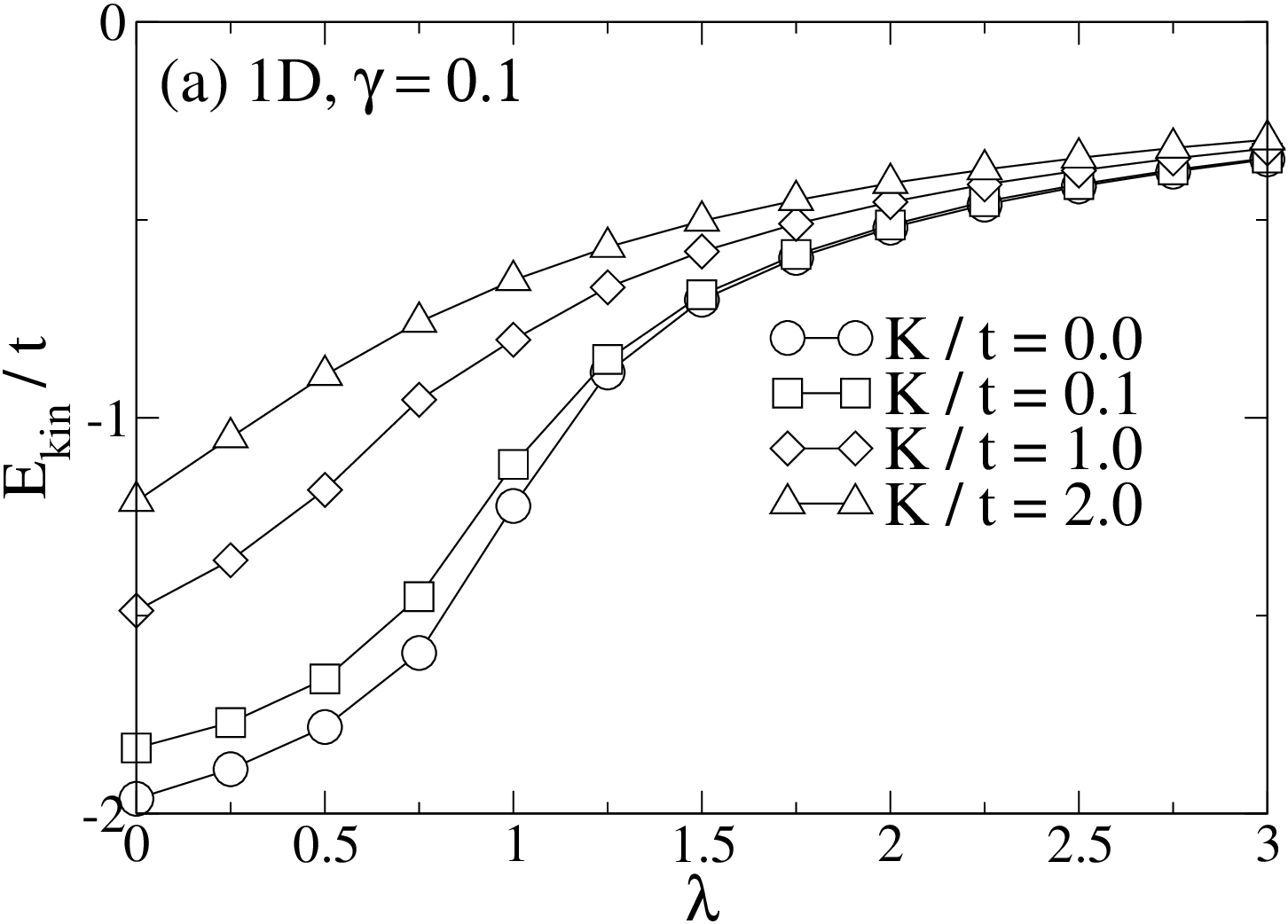}
  \includegraphics[height=0.34\textwidth]{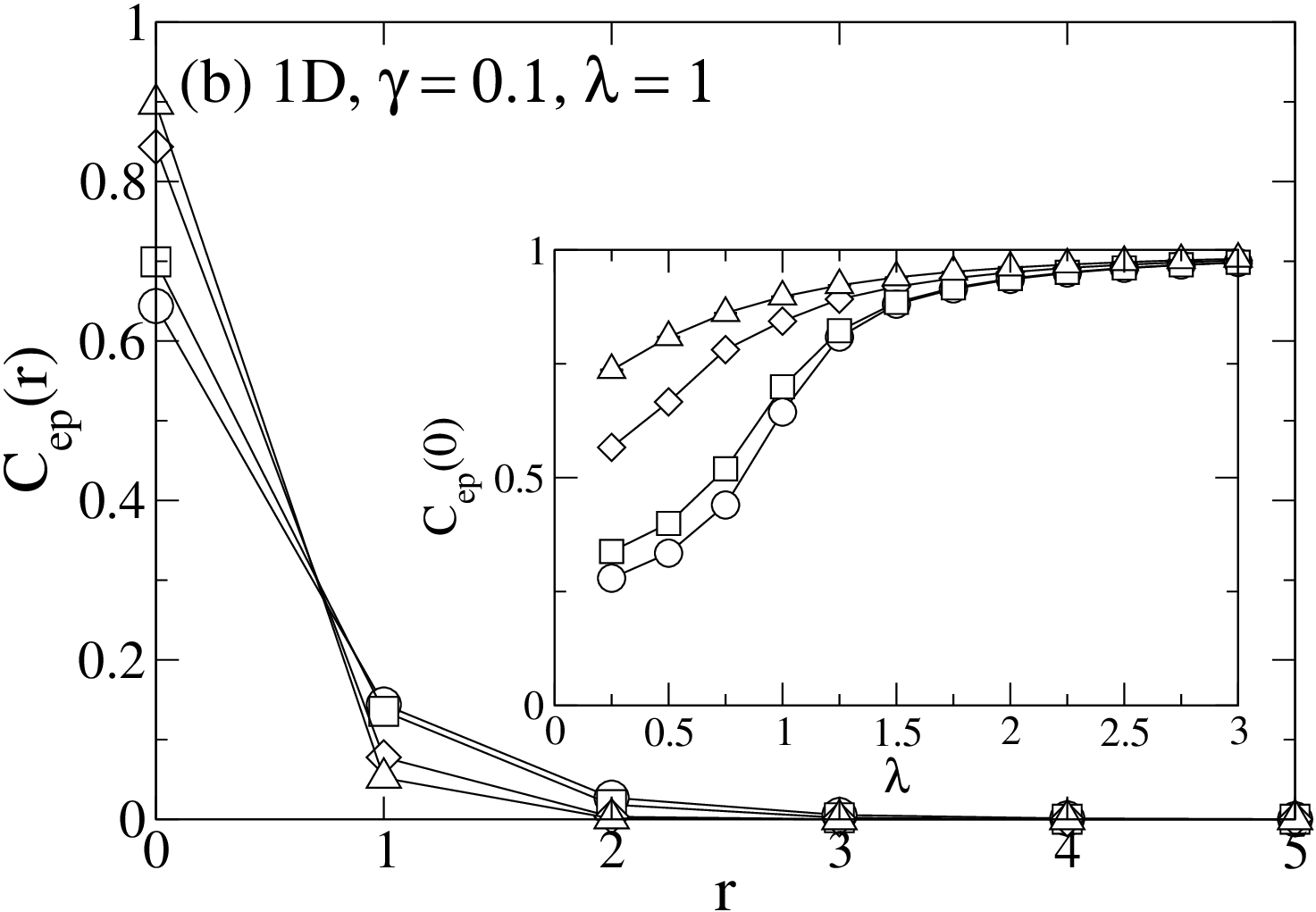}
  \end{center}
  \caption{\label{fig:p_1d_0.1}%
    (a) Kinetic energy $\Ek$ of one electron in a 1D system in the
    adiabatic regime ($\gamma=0.1$) as a
    function of electron-phonon coupling strength $\lambda$ for different values of the
    confinement strength $K$. Here and in subsequent figures $N=31$, $\beta
    t=15$ and $\dtau=0.05$. Lines are guides to the eye only.
    (b) Electron-lattice correlation function $C_\text{ep}(r)$ as a function
    of distance $r$ for the same parameters as in (a) and $\lambda=1$. The
    inset shows $C_\text{ep}(0)$ as a function of $\lambda$.
  }
\end{figure}
\begin{figure}
  \begin{center}
  \includegraphics[height=0.34\textwidth]{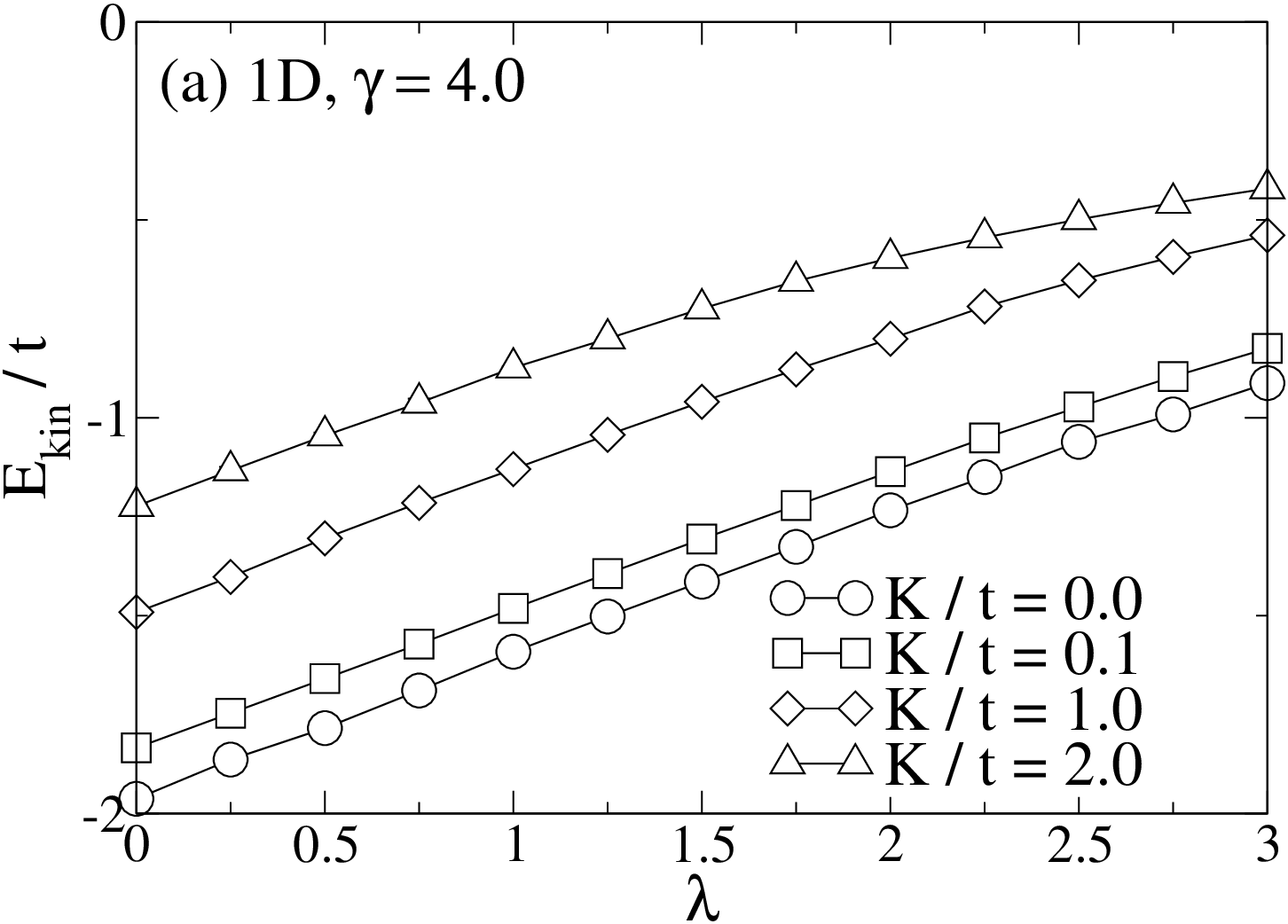}
  \includegraphics[height=0.34\textwidth]{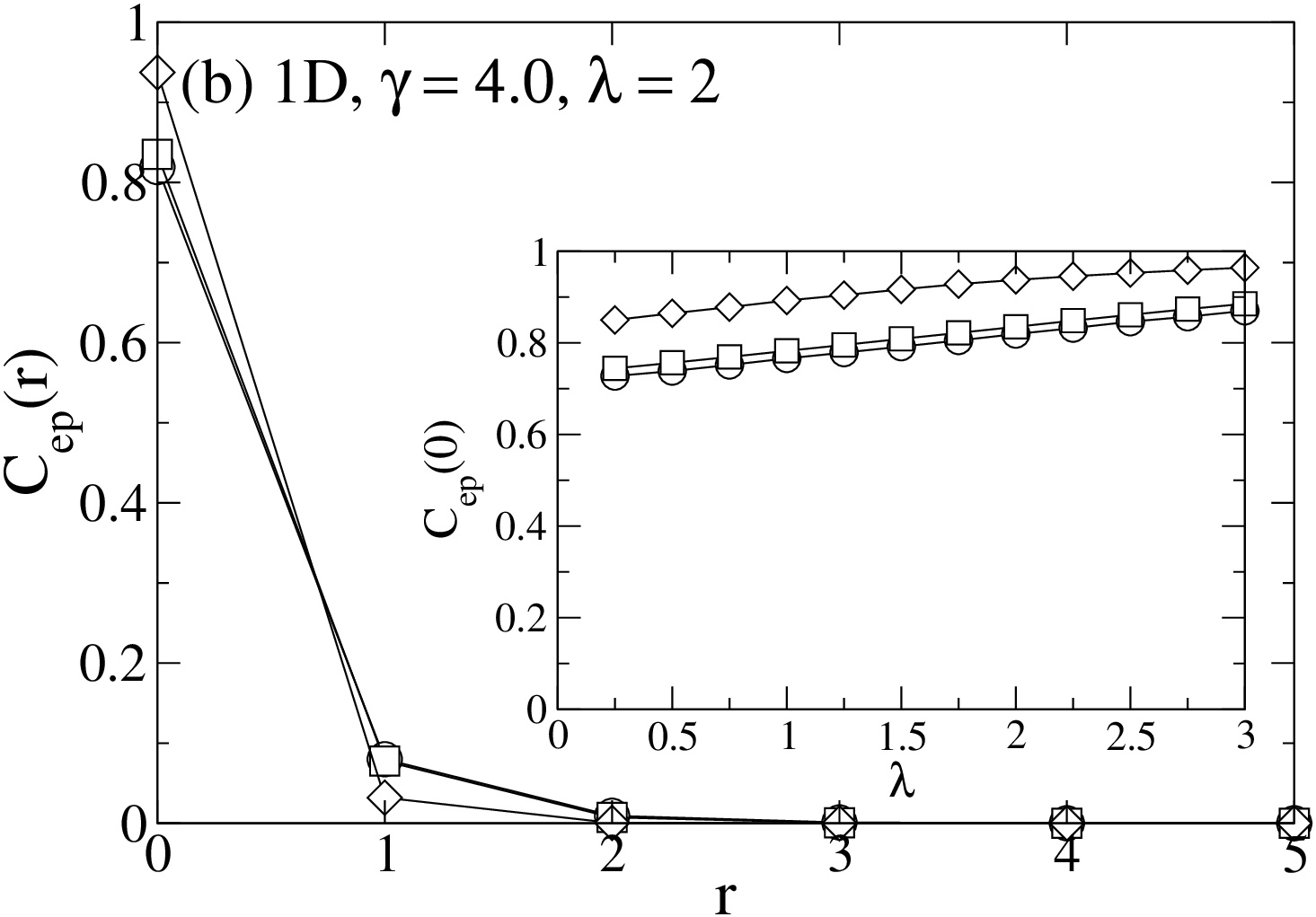}
  \end{center}
  \caption{\label{fig:p_1d_4.0}%
    As in figure~\ref{fig:p_1d_0.1} but for the anti-adiabatic regime ($\gamma=4$).
  }
\end{figure}

Figure~\ref{fig:p_1d_0.1}(a) shows results for a single electron in a 1D
system for $\gamma=0.1$. For $K=0$. starting with the non-interacting value
$-2t$, the absolute value of the kinetic energy is reduced with increasing
$\lambda$.  The crossover near $\lambda=1$ is continuous (the same also
applies in higher dimensions \cite{Loe88}).

Turning on the confinement, we see that results are very similar for small
$K/t=0.1$ (weak confinement). Remarkably, in the case of strong
confinement---the size of the quantum dot is very small for $K/t=2$---the
kinetic energy displays the convex behaviour characteristic of the
strong-coupling regime already near $\lambda=0$. This may be understood as a
result of squeezing of the polaron state: For weak coupling, the free ($K=0$)
polaron size is larger than the dot size set by $K$, so that the kinetic
energy is determined mainly by the confinement. In contrast, for strong
coupling, the polaron size is smaller than the dot size and the kinetic
energy is almost identical to the case $K/t=0$. Therefore, all curves tend to the same
value $-2t I(1)/I(0)$ as $\lambda\to\infty$.

The effect of the confinement on the polaron size at the critical coupling
$\lambda=1$ (for $K=0$) is illustrated by the results for the
electron-lattice correlation function $C_\text{ep}(r)$ in
figure~\ref{fig:p_1d_0.1}(b). For small $K$, the lattice distortion
surrounding the electron extends over about two lattice sites (large
polaron). The effect of confinement is to reduce the polaron size, as is well
visible especially for $K/t=2$.

In the inset of figure~\ref{fig:p_1d_0.1}(b) we present data for
$C_\text{ep}(0)$ as a function of $\lambda$. As pointed out in
\cite{CaCiGr98}, the small-polaron crossover occurs at the point where the
slope of $C_\text{ep}(0)$ has its maximum, which in the present case confirms
the critical coupling $\lambda=1$ obtained from energetic considerations
($K=0$). Furthermore, numerical differentiation of the data shown clearly
indicates that the critical coupling is reduced by about 50 \% in the strong
confinement regime $K/t\geq1$ owing to the smaller non-interacting
($\lambda=0$) kinetic energy.

Figure~\ref{fig:p_1d_4.0} shows results for the anti-adiabatic regime
($\gamma=4$). For $K=0$, the kinetic energy in (a) shows a very smooth
dependence on the electron-phonon coupling strength. The critical value for
the small-polaron crossover is given by $g^2=1$ ($\lambda=2$), but no
pronounced changes in $\Ek$ are visible. For $K>0$, the form of the kinetic
energy curve does not change noticeably, because the free polaron size is
smaller than the confinement length even for $g^2<1$.

The electron-lattice correlation function in figure~\ref{fig:p_1d_4.0}(b)
confirms the smaller size of the polaron at the $K=0$ critical coupling (cf
figure~\ref{fig:p_1d_0.1}(b)). Consequently, the dependence on $K$ is much
weaker than in the adiabatic regime. As for $\gamma=0.1$, the critical
coupling for small-polaron formation is substantially reduced for large
$K/t$, as can be inferred from the slope of $C_\text{ep}(0)$.

\begin{figure}
  \begin{center}
  \includegraphics[height=0.34\textwidth]{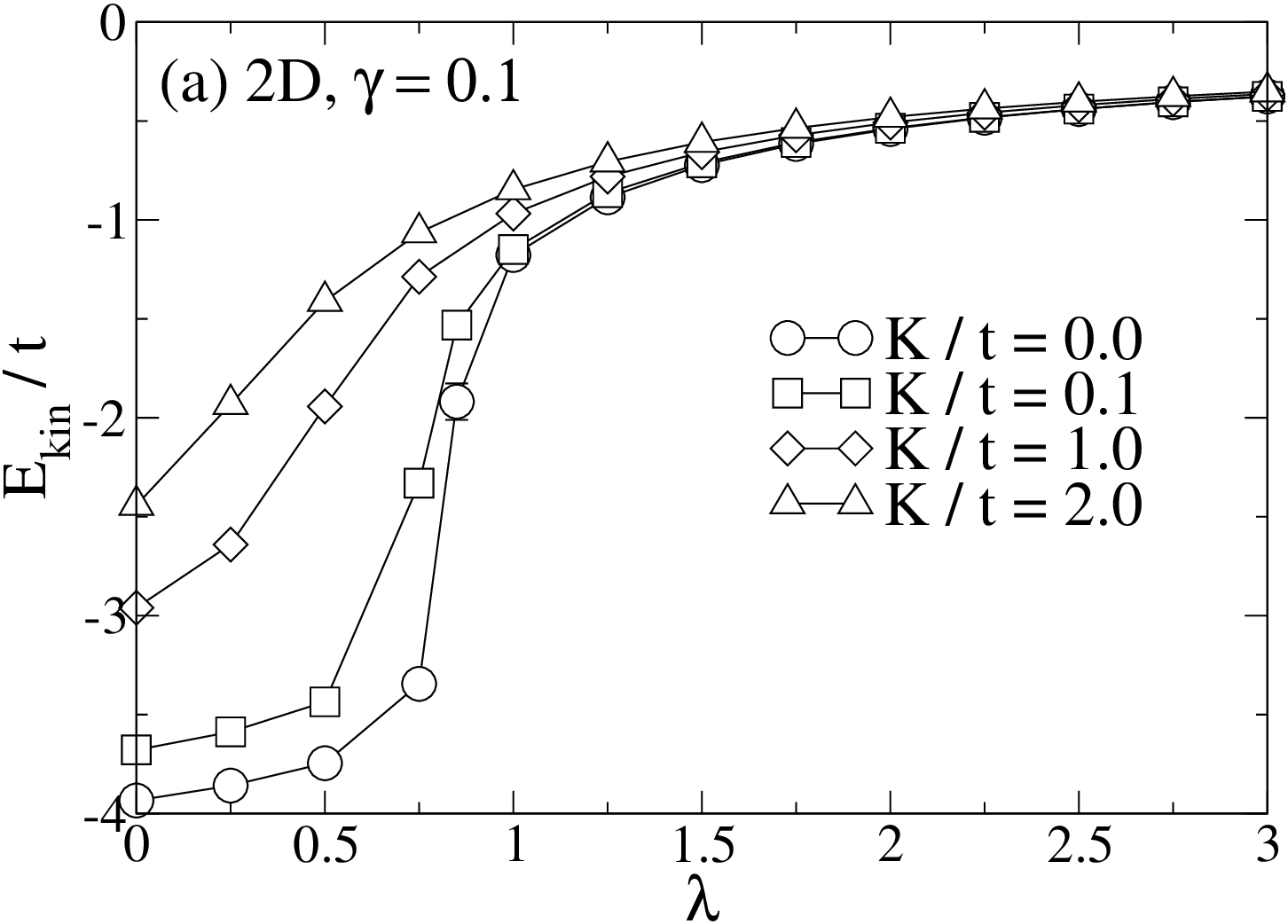}
  \includegraphics[height=0.34\textwidth]{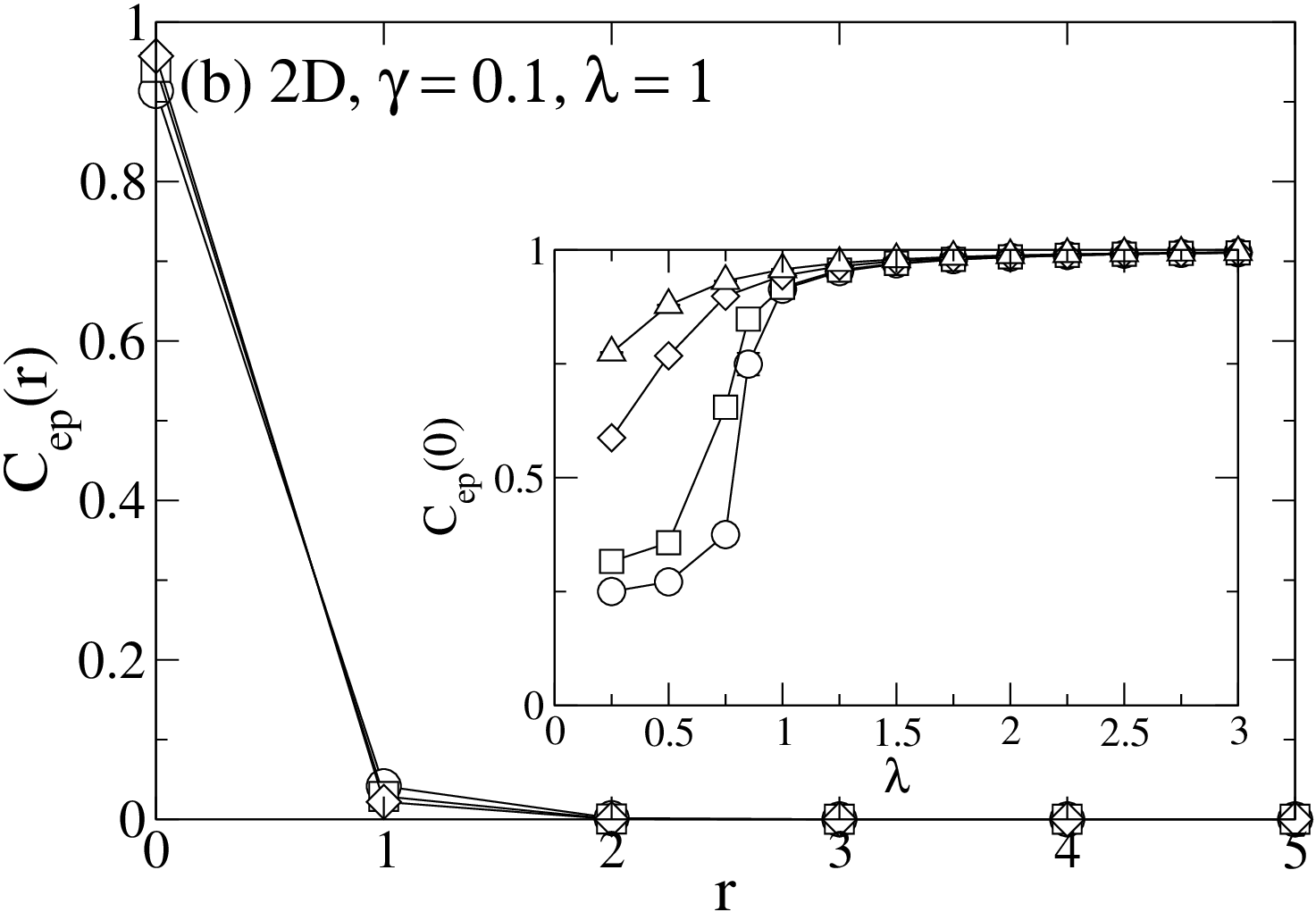}
  \end{center}
  \caption{\label{fig:p_2d}%
    As in figure~\ref{fig:p_1d_0.1} but for a 2D system.
  }
\end{figure}

\subsubsection{Higher dimensions}

For the non-confined case, it is known that the small-polaron crossover
becomes more abrupt as a function of $\lambda$ in higher dimensions. This is
well visible from the 2D results for $\Ek$ shown in figure~\ref{fig:p_2d}(a),
as well as from the inset of figure~\ref{fig:p_2d}(b).

The effect of confinement is again to reduce the kinetic energy. For
$K/t=2$, the convex strong-coupling behaviour is seen similar to $\rD=1$. An
important difference to the 1D case is that the curves for different $K$
become practically indistinguishable for $\lambda>2$, whereas in 1D
(figure~\ref{fig:p_1d_0.1}(a)) such a collapse occurs for $\lambda\gtrsim3$.
The origin of this effect is revealed in figure~\ref{fig:p_2d}(b). Obviously,
the polaron is smaller in 2D than in 1D for the same $\lambda$, reducing the
effect of $K$. The inset again reveals a reduction of the critical coupling
with increasing confinement. Similar behaviour is also found in three
dimensions (not shown), with an even stronger dependence of $\Ek$ and
$C_\text{ep}(r)$ on $\lambda$.

Let us discuss the relation of our results to previous work. The common
conclusion is that the confinement demobilises the carrier, which manifests
itself either in an increase of effective mass, or a decrease of kinetic
energy and the polaron size \cite{YiEr91,YiEr91_2,Sa96,PoFoDeBaKl99,ChMu01}.
Also in accordance with existing work, we find that the polaronic correction
to the total energy is larger for larger $K$. For example, $\Delta
E^\text{f}=[E^\text{f}_{\lambda=0}-E^\text{f}_{\lambda=1}]/E^\text{f}_{\lambda=0}\approx0.54$
for $K/t=0.1$, whereas $\Delta E^\text{f}\approx1.08$ for $K/t=2$.

The reduction of the critical coupling for the existence of a small polaron
due to confinement complies with the growth of the strong-coupling region at
the expense of the weak-coupling region in the phase diagram pointed out in
\cite{PoFoDeBaKl99}.

Furthermore, we observe a similar dependence of the influence of confinement
on dimensionality as for the continuum model \cite{Sa96} (in this work the
effective mass was considered). Comparing the values of the kinetic energy at
$\lambda=1$, we find for the ratio $\Ek(K)/\Ek(K=0)=0.91$
($K/t=0.1$), 0.66 ($K/t=1$), and 0.53 ($K/t=2$) in 1D, and 0.98, 0.82, and
0.72 in 2D. Hence, confinement effects are significantly
stronger in 1D than in 2D, and this trend extends to $\rD=3$.

\subsection{Bipolaron}

We now come to the case of two electrons of opposite spin, which can form a
bound bipolaron state given sufficiently strong electron-phonon interaction.
Owing to the Coulomb repulsion, the physics becomes much richer, and we again
begin our discussion with the one-dimensional case.

Compared to the one-electron problem, significantly less work has been done
on the bipolaron problem in the framework of the Holstein-Hubbard model.
Nevertheless, from existing studies (see
\cite{AlMo95,HoAivdL04,Mac04,HovdL05} and references therein), the basic
physics is understood. A summary of (approximate) conditions for the
existence of the different ground states is given in
table~\ref{tab:bipolaron:bipolaronconditions}.

\begin{table}[width=0.55\textwidth,ht]
  \centering
  \caption{\label{tab:bipolaron:bipolaronconditions}Conditions for the
    existence of different singlet bipolaron states in the
    one-dimensional Holstein-Hubbard model at weak (WC) and strong
    electron-phonon coupling (SC) \cite{BoKaTr00,HovdL05}.}
  \vspace*{1em}
  \begin{tabular}{c|c}\hline\hline
    \multicolumn{2}{c}{$U=0$}\\\hline
    Large           & Small          \\
    bipolaron       & bipolaron      \\\hline
    $\lambda<0.5$ ($\gamma<1$)   & $\lambda>0.5$ ($\gamma<1$) \\
    or              & and            \\
    $g<0.5$       ($\gamma>1$) & $g>0.5$  ($\gamma>1$)      \\\hline
    \hline
  \end{tabular}\hspace*{0.5em}%
  \begin{tabular}{c|c|c}\hline\hline
    \multicolumn{3}{c}{$U>0$} \\\hline
    Two     & Inter-site & Small    \\
    polarons& bipolaron  & bipolaron\\\hline
    $U>2\Ep$ (WC) & $U<2\Ep$ (WC)  &            \\
                  &                &  $U\ll2\Ep$\\
    $U>4\Ep$ (SC) & $U<4\Ep$ (SC)  &            \\\hline
    \hline
  \end{tabular}
\end{table}

\subsubsection{One dimension}

\begin{figure}
  \begin{center}
  \includegraphics[height=0.34\textwidth]{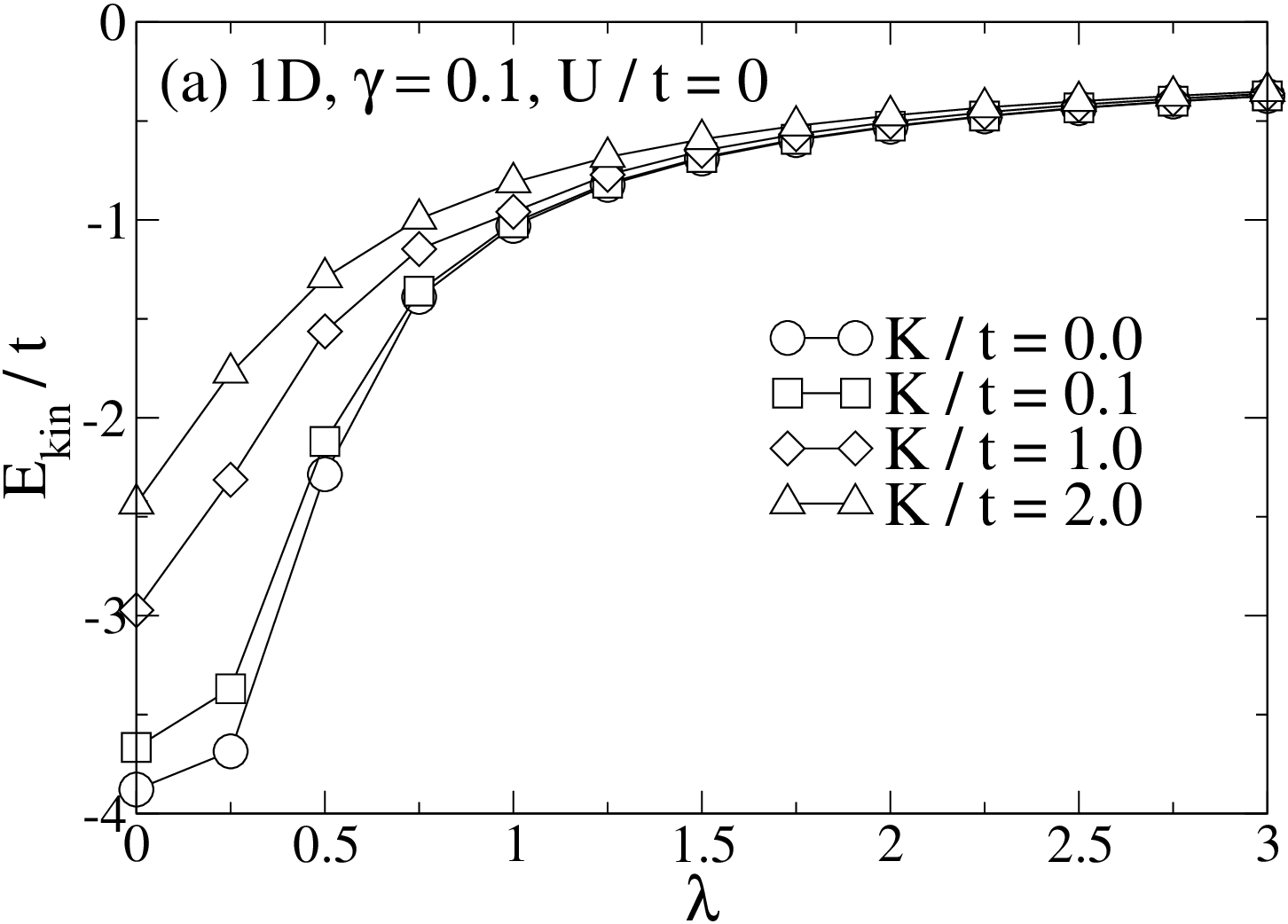}
  \includegraphics[height=0.34\textwidth]{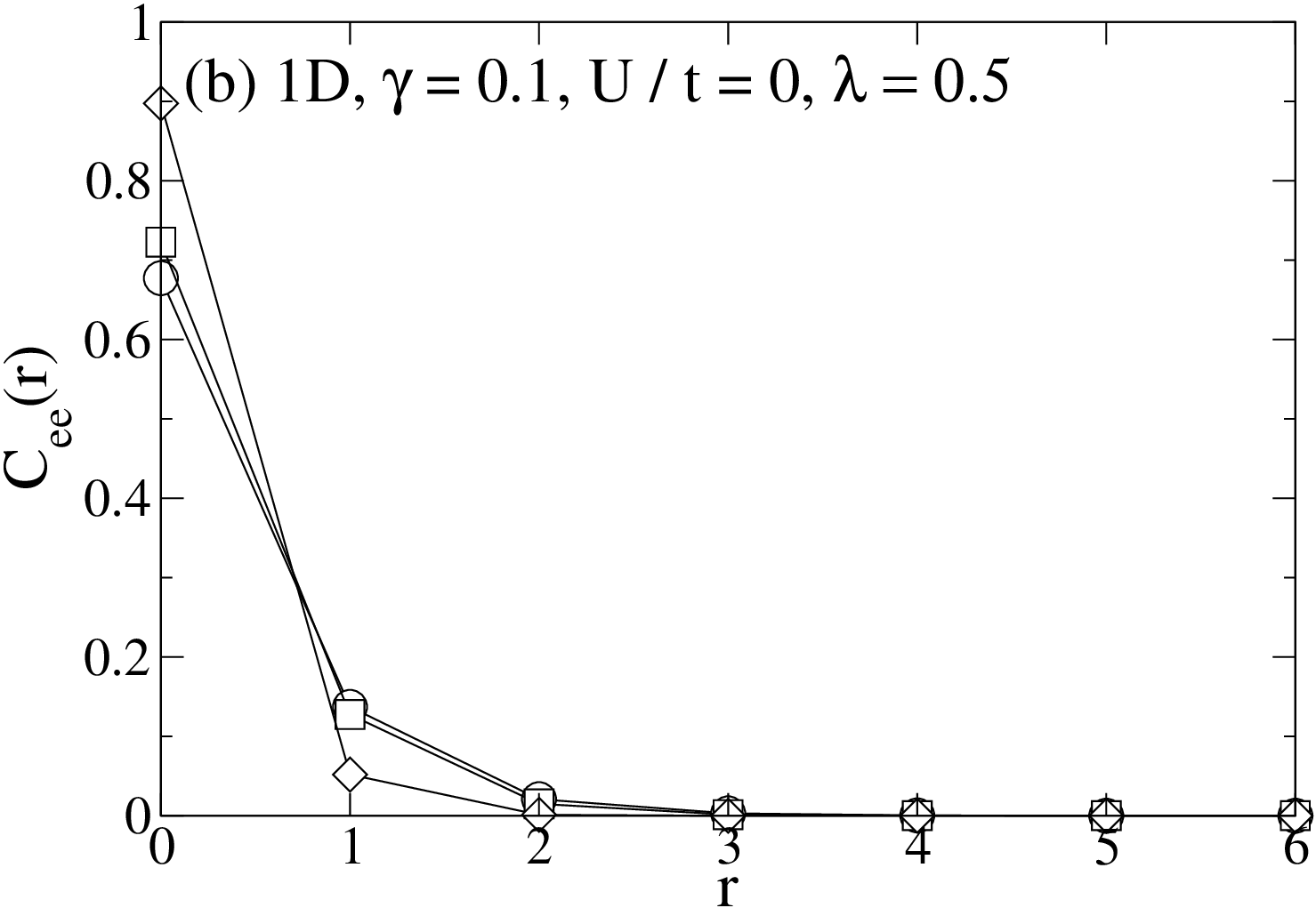}
  \end{center}
  \caption{\label{fig:bp_1d_U0}%
    (a) Kinetic energy $\Ek$ of two electrons in a 1D system in the
    adiabatic regime ($\gamma=0.1$) for $U=0$ as a function of
    electron-phonon coupling strength
    $\lambda$ for different values of the confinement strength $K$. 
    (b) Electron-electron correlation function $C_\text{ee}(r)$ as a function
    of distance $r$ for the same parameters as in (a) and $\lambda=0.5$.  
  }
\end{figure}

\begin{figure}
  \begin{center}
  \includegraphics[height=0.34\textwidth]{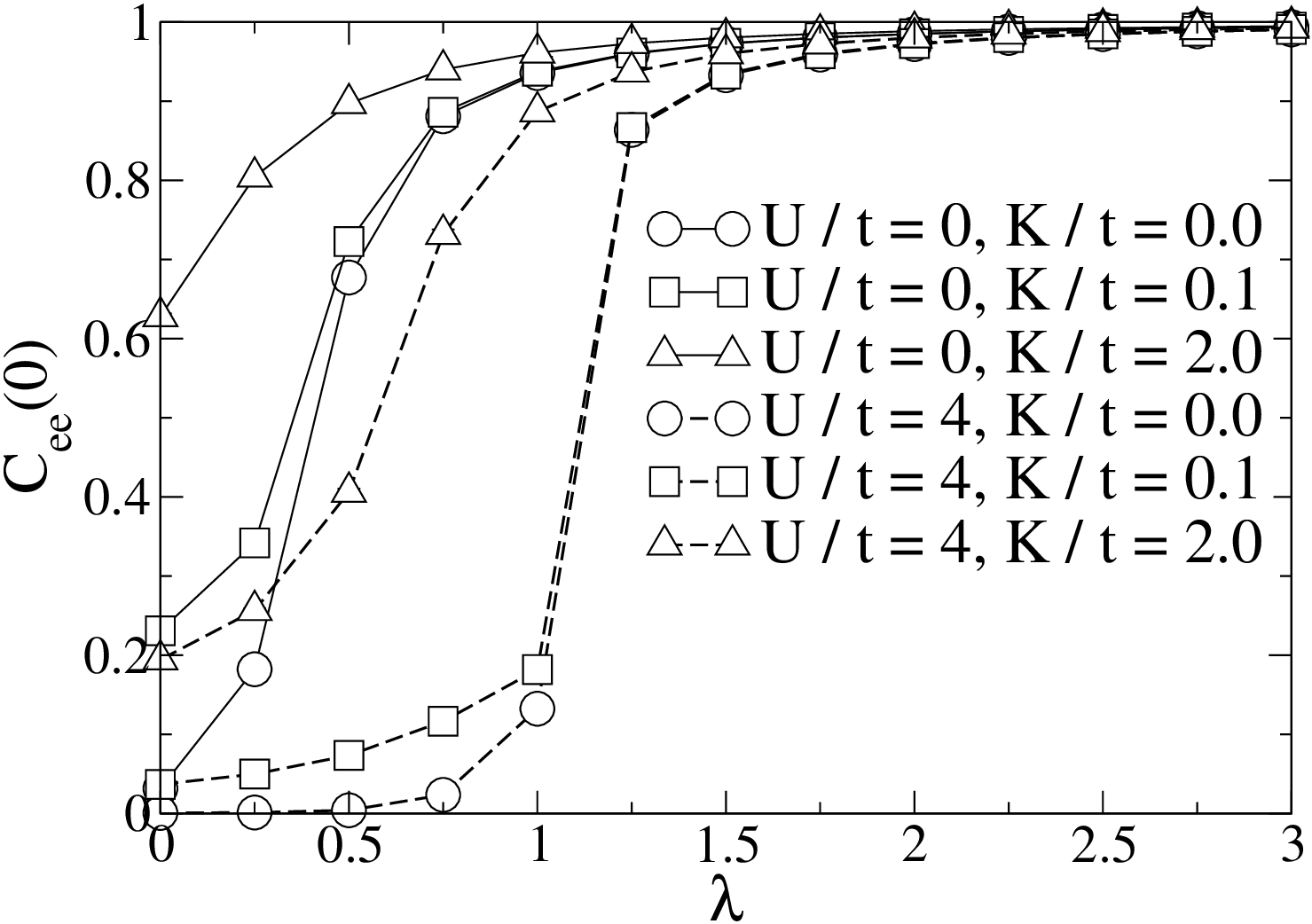}
  \end{center}
  \caption{\label{fig:bp_1d_Cee}%
    Electron-electron correlation function $C_\text{ee}(0)$ as a function of
    electron-phonon coupling strength $\lambda$ for a 1D system, $\gamma=0.1$,
    and different values of the Hubbard
    repulsion $U$ and the confinement strength $K$.
  }
\end{figure}

Figure~\ref{fig:bp_1d_U0} shows results for the simplest case $U=0$, in the
adiabatic regime ($\gamma=0.1$). For $K=0$, the two electrons form a bound
state if $4\Ep>4t$ ($\lambda>0.5$), where $4\Ep$ is the bipolaron binding
energy in the atomic limit and $4t$ is the free kinetic energy of the two
electrons. The corresponding crossover from a large bipolaron (no unbound
polarons exist for $U=0$) to a small (on-site) bipolaron leads to a
noticeable decrease of the (absolute) kinetic energy in
figure~\ref{fig:bp_1d_U0}(a). As for the polaron, the critical value for the
crossover may be determined from the slope of $C_\text{ep}(0)$ or
$C_\text{ee}(0)$ (see figure~\ref{fig:bp_1d_Cee}).

Apart from enforcing polaron effects, confinement enhances the probability
for the two electrons to occupy the same region of the system (the vicinity
of the centre of the harmonic potential). This in turn leads to a stronger
pairing tendency. Any $K/t>0$ decreases the kinetic energy in the weak- and
intermediate coupling regime. For strong confinement, the kinetic energy
exhibits the typical strong-coupling dependence on $\lambda$. The reduction
of the bipolaron size (the average distance between the two carriers) due to
confinement is shown in figure~\ref{fig:bp_1d_U0}(b) for the free critical
coupling $\lambda=0.5$. Finally, figure~\ref{fig:bp_1d_Cee} reveals that the
critical coupling for small-bipolaron formation is substantially smaller in a
strongly confined system.

It is more realistic to consider the case of finite Coulomb repulsion
$U/t=4$.  For $K=0$, strong-coupling theory predicts a ground state with two
unbound (large) polarons for $2\Ep\lesssim U$. In the vicinity of $2\Ep=U$,
the so-called inter-site bipolaron exists, with the electrons being most
likely to occupy neighbouring lattice sites, and with only slightly enhanced
effective mass as compared to two unbound polarons
\cite{BoKaTr00,WeFeWeBi00,HovdL05}. The origin of this state are non-local
exchange interaction processes \cite{BoKaTr00} which allow the particles to
gain potential energy from the coupling to phonons, and at the same time
avoid the penalty due to Hubbard repulsion.

\begin{figure}
  \begin{center}
  \includegraphics[height=0.34\textwidth]{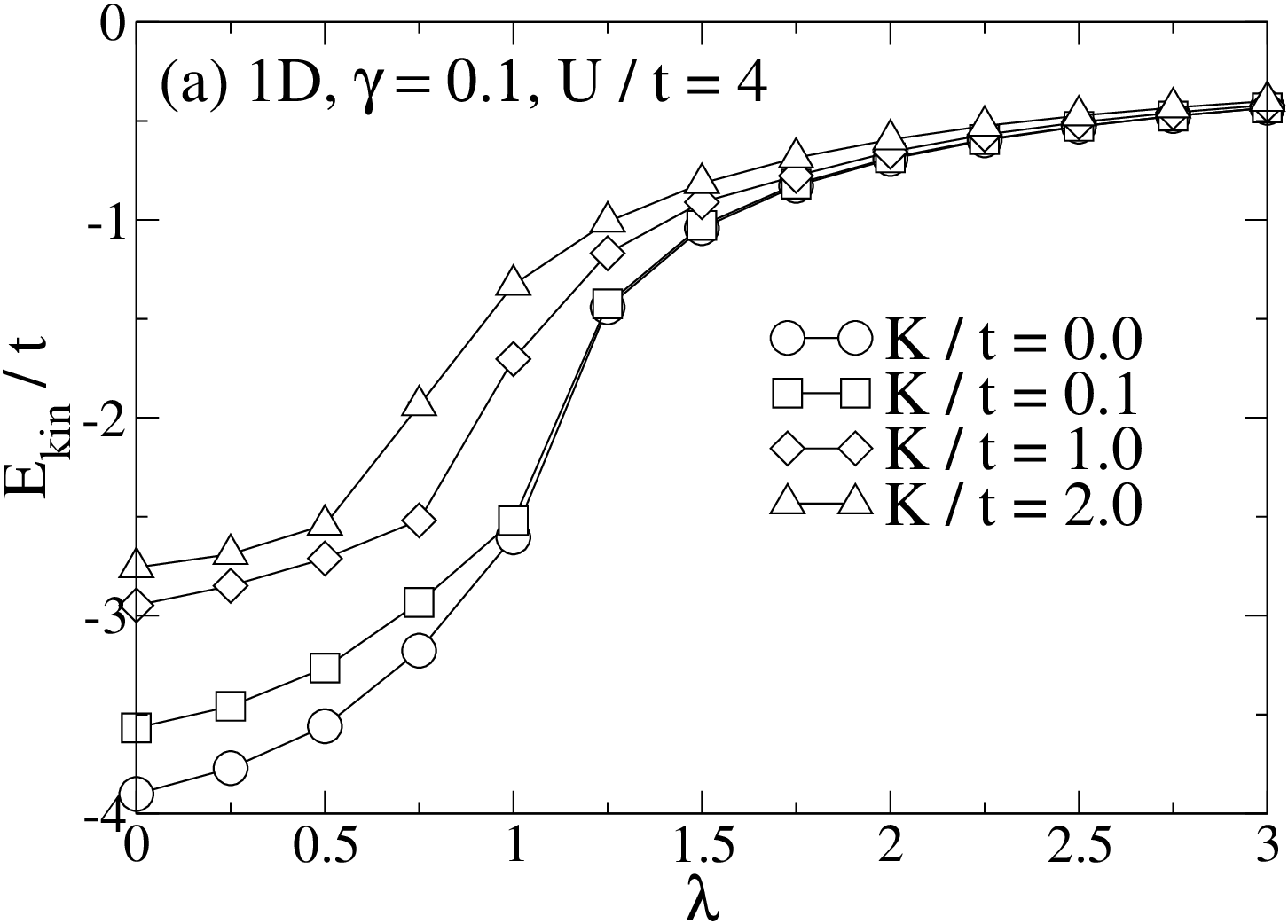}
  \includegraphics[height=0.34\textwidth]{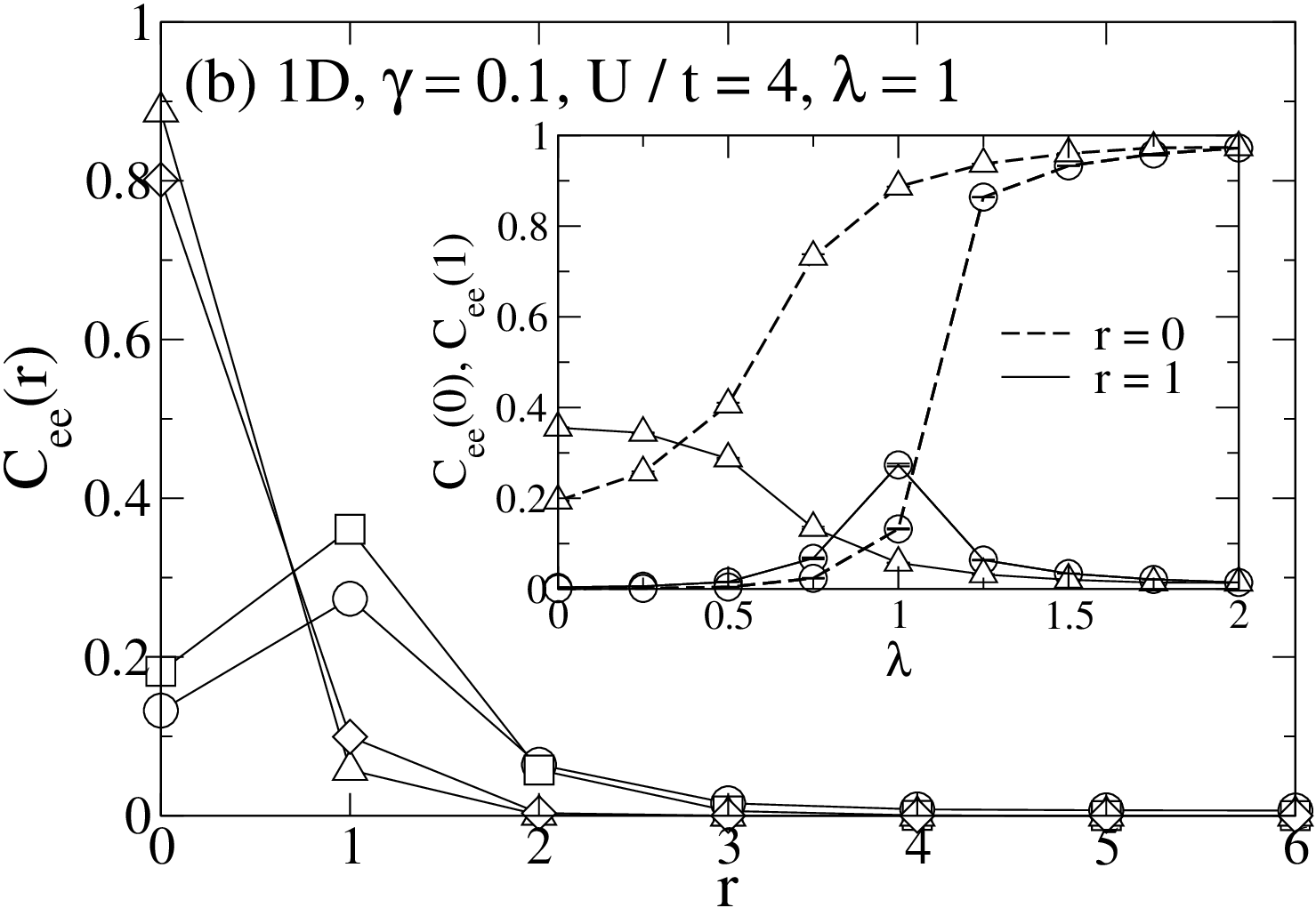}
  \end{center}
  \caption{\label{fig:bp_1d_U4}%
    As in figure~\ref{fig:bp_1d_U0} but for $U/t=4$. The inset in (b) shows
    the electron-electron correlation functions
    $C_\text{ee}(0),\,C_\text{ee}(1)$.
  }
\end{figure}

Remarkably, figure~\ref{fig:bp_1d_U4}(a) reveals a critical coupling for the
small-bipolaron crossover of $2\Ep\approx U$ ($\lambda\approx1$, see also
inset of figure~\ref{fig:bp_1d_U4}(b)), which is in a surprisingly good
agreement with the anti-adiabatic strong-coupling condition despite
$\gamma=0.1$. Our findings extend previous calculations limited either to
$\gamma=1$ \cite{BoKaTr00} or small systems \cite{HovdL05}, and a phase
diagram for the important adiabatic regime will be presented elsewhere.

As for $U=0$, confinement leads to a reduction of the kinetic energy and the
average distance between the two electrons (figure~\ref{fig:bp_1d_U4}(b)).
However, owing to the on-site repulsion, the results for $\Ek$ exhibit weak
and strong-coupling behaviour even for $K/t=2$.

An interesting point concerns the effect of confinement on the inter-site
state. The inset in figure~\ref{fig:bp_1d_U4}(b) shows the electron-electron
correlation functions $C_\text{ee}(0)$ and $C_\text{ee}(1)$. For $K=0$, in
the vicinity of $\lambda=1$, $C_\text{ee}(1)>C_\text{ee}(0)$ as
characteristic for the inter-site bipolaron. In contrast, for strong
confinement $K/t=2$, the region of existence of the latter is limited to
small $\lambda$. Clearly, for sufficiently strong confinement, the inter-site
state will disappear.

In previous strong-coupling calculations in the framework of the continuum
model \cite{MuCh96,ChMu01,SeEr00}, it has been found that the bound bipolaron
state becomes unstable at very strong confinement, which was attributed to
the increase of the Coulomb interaction energy for two spatially close
particles. However, this effect has been argued \cite{PoFoDeBaKl99,SeEr00} to
be due to the approximations made, and path-integral calculations
\cite{PoFoDeBaKl99} as well as analytical and QMC calculations
\cite{WaOrPh97} suggest that a bound state also exists at weak coupling in a
confined system.

Preliminary calculations of the bipolaron binding energy
$E^\text{f}(N_\text{e}=2)-2E^\text{f}(N_\text{e}=1)$ for $\gamma=0.1$ and
$U/t=4$ at $K/t=0.1$ and $K/t=2.0$ in one dimension reveal that a weakly
bound bipolaron can indeed become unbound due to confinement. This effect of
on-site Coulomb interaction is expected to be most pronounced in one
dimension, and this issue will be further investigated in future work.

It is worth mentioning the relation of our calculations to recent work on
molecular quantum dots \cite{AlBr03,KoRavO06} based on models with a
single molecular level and a vibrational mode, weakly coupled to metallic
leads. In such systems, the bipolaron binding energy can compensate for the
Hubbard repulsion $U$, giving rise to a net attraction and thereby favouring pair
tunneling of electrons. Due to the absence of a sizable hopping amplitude,
the ground-state properties of such a single-site molecule are determined
mainly by the atomic-limit physics of the present model. In particular, the
inter-site bipolaron state will be strongly suppressed, and the
strong-coupling criteria of table~\ref{tab:bipolaron:bipolaronconditions} are
expected to hold. The situation may be expected to be different for
finite-size molecules.

\subsubsection{Higher dimensions}

Numerical results for the unconfined Holstein-Hubbard model in $\rD>1$ have
been reported before in 2D for $\gamma=1$ \cite{Mac04}, and in 3D for
classical phonons \cite{deRaLa86}, but no reliable results are available for
the important adiabatic regime $\gamma\ll1$. We restrict the discussion to
the more general case $U>0$.

\begin{figure}
  \begin{center}
    \includegraphics[height=0.34\textwidth]{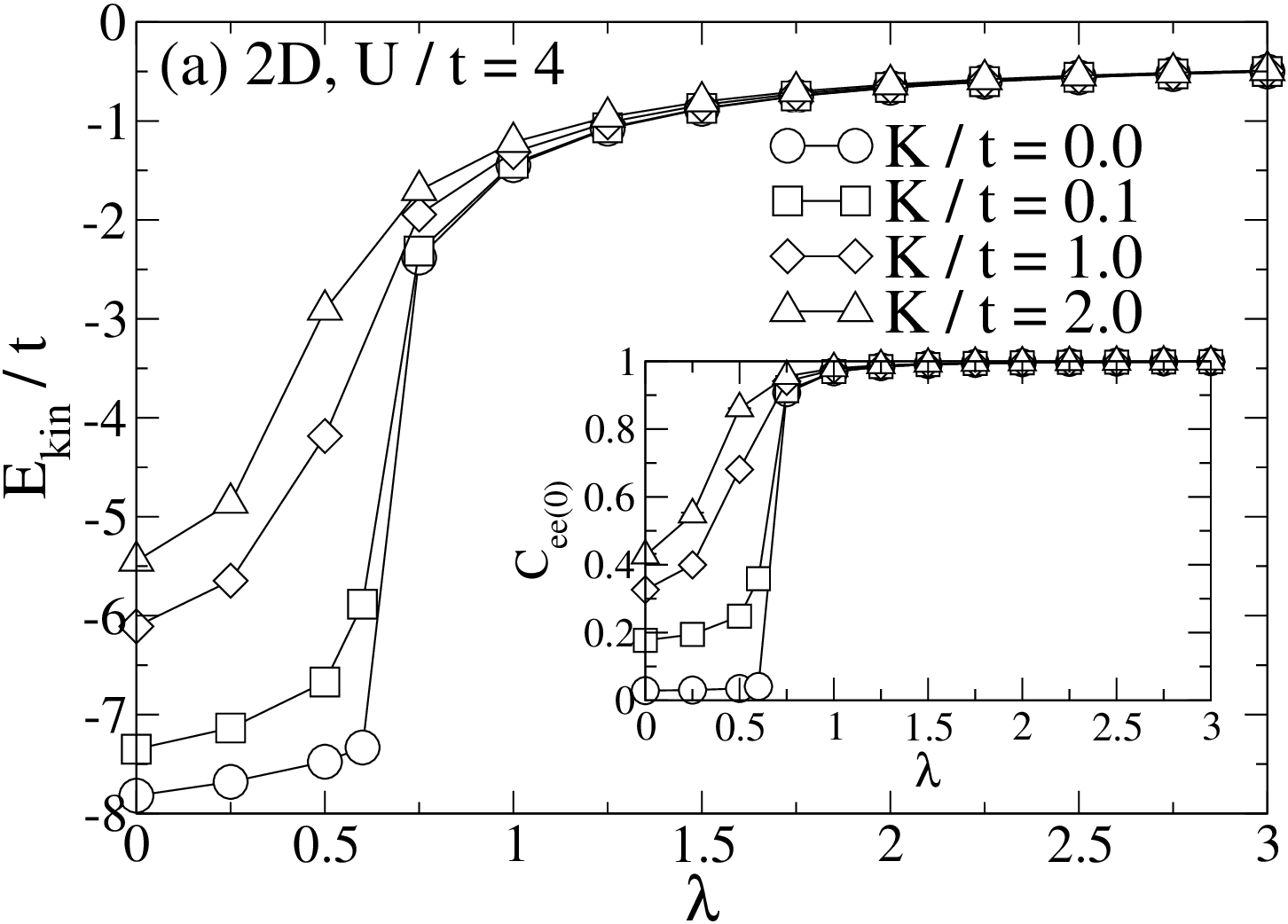}
    \includegraphics[height=0.34\textwidth]{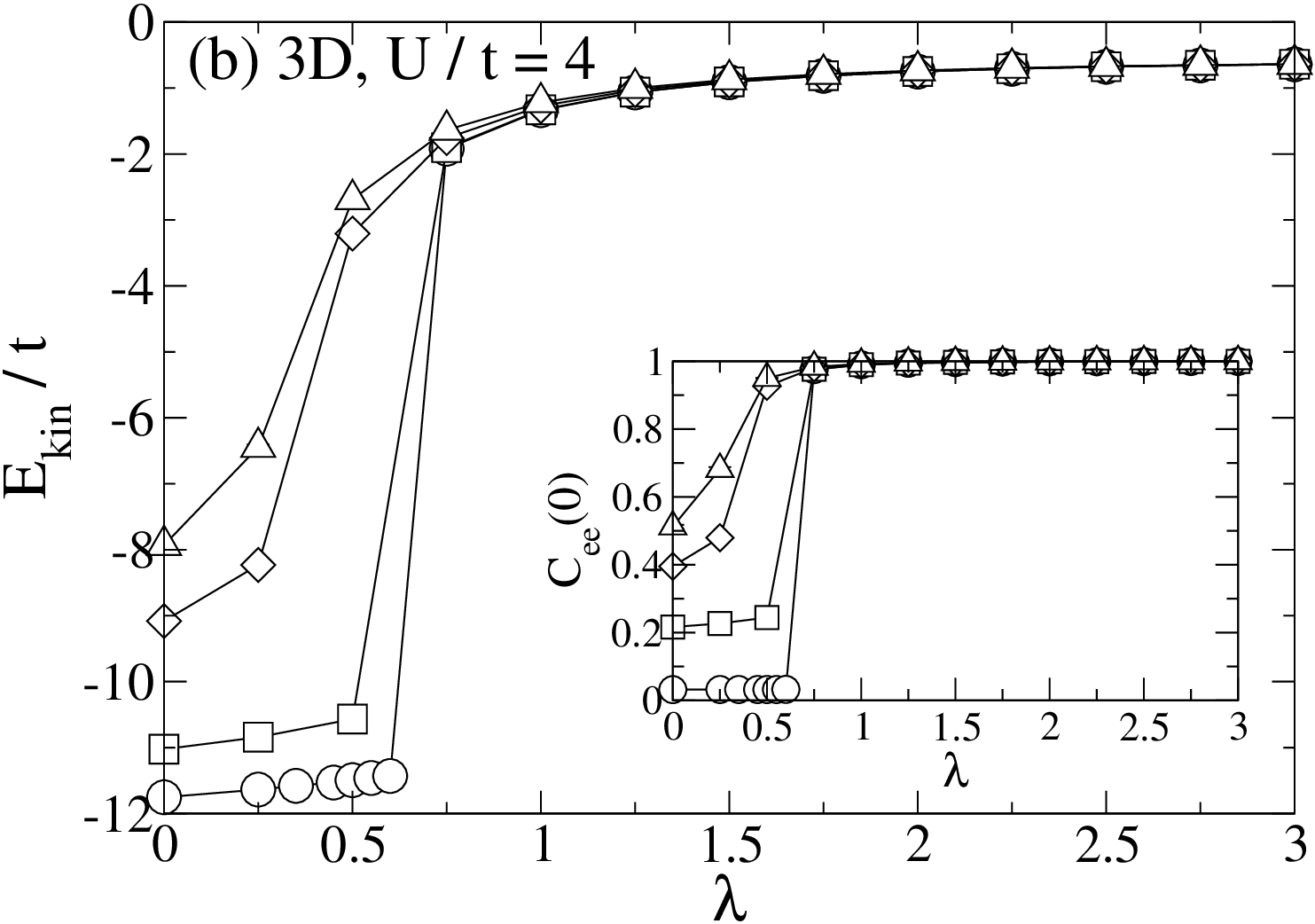}
  \end{center}
  \caption{\label{fig:bp_23d}%
    Kinetic energy $\Ek$ for two electrons in the adiabatic regime
    ($\gamma=0.1$) for $U/t=4$ as a function of coupling strength $\lambda$
    for different values of the confinement strength $K$ in (a) two
    dimensions, (b) three dimensions.  Insets show the electron-electron
    correlation function $C_\text{ee}(r=0)$ versus $\lambda$.}
\end{figure}

As in the polaron problem, the crossover to a small bipolaron becomes more
pronounced in higher dimensions (note the different ordinate scales in
figures~\ref{fig:bp_1d_U4}(a) and \ref{fig:bp_23d}(a) and (b)). The critical
coupling for $K=0$ from the strong-coupling approximation is again set by
$U\approx2\Ep$, corresponding to $U/W=\lambda$. Indeed, for $\rD=2$
(figure~\ref{fig:bp_23d}(a)), we find the crossover near $\lambda=0.5$ in
agreement with this condition. However, in three dimensions, the crossover
occurs slightly above $\lambda=0.5$, whereas the strong-coupling result
yields $\lambda=0.33$. This deviation may be explained by the larger number
of nearest-neighbour sites in 3D, which may enhance the stability of extended
(inter-site) bipolaron states.

Confinement again leads to a reduction of the critical coupling, as well as
to an enhancement of bipolaron effects. As for the a single electron, we can
calculate the change of the kinetic energy with confinement (relative to
$K=0$) at $\lambda=1$. This yields 0.89 in two dimensions, and 0.95 in three
dimensions, so that we can conclude that confinement effects are again more
noticeable in lower dimensions.

%

%
%
%
%%%%%%%%%%%%%%%%%%%%%%%%%%%%%%%%%%%%%%%%%%%%%%%%%%%%%%%%%%%%%%%%%%%%%
\section{Summary and outlook}\label{sec:summary}
%%%%%%%%%%%%%%%%%%%%%%%%%%%%%%%%%%%%%%%%%%%%%%%%%%%%%%%%%%%%%%%%%%%%%
%
%
%

Unbiased quantum Monte Carlo studies of the Holstein-Hubbard model with
additional harmonic confinement on a discrete lattice have been carried out
in one to three dimensions, considering the cases of one and two electrons.
Technical difficulties encountered in simulations using the present
world-line method, some of which were overlooked in previous work, have been
revealed.

The effect of confinement on the formation of polarons and bipolarons has
been investigated. Despite considerable simplifications inherent to the
model, we believe that our exact numerical results are relevant for quantum
dot systems.

In addition to providing unbiased results for all physically relevant
parameter regimes, we have for the first time reported on results for a
(multi-site) cluster model of a quantum dot. The latter is particularly
important for the very small dots which can be fabricated today, as well as
for a description of the small-(bi-)polaron crossover associated to a local
lattice instability. Quantum phonon effects and electron-electron interaction
were fully taken into account.

For one electron, we find that the basic effect of confinement consists in
enhancing polaronic effects---the polaron size and the critical coupling for
the existence of a small-polaron both decrease. The influence of confinement
is largest in a one-dimensional system in the adiabatic regime, and becomes
significantly smaller with increasing dimension or phonon frequency. In the
strongly confined regime, the polaron state is squeezed, giving rise to
small-polaron physics even for weak or intermediate coupling.

We have presented the first accurate results for the unconfined
Holstein-Hubbard bipolaron in the adiabatic regime in more than one
dimension. We find indications that confinement can counteract bipolaron
formation in the presence of Coulomb repulsion, but this issue needs further
investigation also in the framework of a model with long-range interactions.
For finite on-site repulsion, weak-coupling behaviour survives even for strong
confinement.

Existing work on (bi-)polarons in quantum dots is almost exclusively
restricted to equilibrium properties, although most technical applications
fall into the non-equilibrium regime. Since the fundamental effects of
confinement on the ground state are now rather well understood, it is
desirable to consider more general situations in the future. Apart from
effects due to long-range interactions and anharmonic confinement in cluster
models, this includes many-polaron effects \cite{HoHaWeFe06}, coupling of
electrons or excitons to light \cite{PrGrFedVGuTePoLe06}, as well as
transport properties and time-dependent phenomena. Of course, suitable
numerical methods will have to be developed to address these problems, and
work along these lines is in progress.

%%%%%%%%%%%%%%%%%%%%%%%%%%%%%%%%%%%%%%%%%%%%%%%%%%%%%%%%%%%%%%%%%%%%%
\ack

We gratefully acknowledge financial support by the Austrian Science Fund
(FWF) through the Erwin-Schr\"odinger Grant No~J2583, the Deutsche
Forschungsgemeinschaft through SFB 652, KONWIHR, and the European
Science Foundation. We thank H~de Raedt for valuable correspondence, and
G~Hager and G~Wellein for useful discussion. Furthermore, we acknowledge
generous computing time at the TU Graz and the Computing Centre Erlangen.

%%%%%%%%%%%%%%%%%%%%%%%%%%%%%%%%%%%%%%%%%%%%%%%%%%%%%%%%%%%%%%%%%%%%%

%%%%%%%%%%%%%%%%%%%%%%%%%%%%%%%%%%%%%%%%%%%%%%%%%%%%%%%%%%%%%%%%%%%%%%%%%%%%%
%%%%%%%%%%%%%%%%%%%%%       BIBLIOGRAPHY              %%%%%%%%%%%%%%%%%%%%%%%
%%%%%%%%%%%%%%%%%%%%%%%%%%%%%%%%%%%%%%%%%%%%%%%%%%%%%%%%%%%%%%%%%%%%%%%%%%%%%

\section*{References}

%\bibliography{../../book_heraeus/bibliography}
%\bibliographystyle{prsty}

\end{document}